\documentclass{emulateapj}
\usepackage{apjfonts}
\usepackage{hyperref}
\usepackage{amsmath}
\usepackage{float}
\usepackage{threeparttable}
\usepackage{color}

\newcommand{{\HST}}{\emph{HST}}
\newcommand{{\M}}{log($M_{*}$/M$_{\odot}$)}
\newcommand{{\kpc}}{~kpc}
\newcommand{{\n}}{S\'{e}rsic}
\newcommand{\ha}{Our results suggest,}

\graphicspath{{/Users/rebeccaallen/Dropbox/Rebecca/Proj1/paper/resub/fig2/}} 
\slugcomment{ApJ: accepted, 11 April, 2015} 
\shortauthors{\sc Allen et~al.}

\begin{document}

\title{The differential size growth of field and cluster galaxies at $z=2.1$ using the \\ ZFOURGE survey\altaffilmark{$\star$}}
\email{rallen@astro.swin.edu.au}

\author{Rebecca J. Allen\altaffilmark{1,2},
Glenn G. Kacprzak\altaffilmark{1},
Lee R. Spitler\altaffilmark{2,3},
Karl Glazebrook\altaffilmark{1},
Ivo Labb\'e\altaffilmark{5},
Kim-Vy H. Tran\altaffilmark{4},
Caroline M. S. Straatman\altaffilmark{5},
Themiya Nanayakkara\altaffilmark{1},
Ryan F. Quadri\altaffilmark{4},
Michael Cowley\altaffilmark{3},
Andy~Monson\altaffilmark{6},
Casey Papovich\altaffilmark{4},
S.~Eric~Persson\altaffilmark{6},
Glen Rees\altaffilmark{3},
V. Tilvi\altaffilmark{4},
Adam R. Tomczak\altaffilmark{4}
}
                                                                                
\altaffiltext{1}{Swinburne University of Technology, Victoria 3122,
Australia}
\altaffiltext{2}{Australian Astronomical Observatories, PO Box 915, 
North Ryde, NSW 1670, Australia}
\altaffiltext{3}{Department of Physics \& Astronomy, Macquarie University, Sydney, NSW 2109, Australia}
\altaffiltext{4}{George P. and Cynthia Woods Mitchell Institute for Fundamental Physics and Astronomy, and Department of Physics and Astronomy, Texas A\&M University, College Station, TX 77843}
\altaffiltext{5}{Sterrewacht Leiden, Leiden University, NL-2300 RA 
Leiden, The Netherlands}
\altaffiltext{6}{Carnegie Observatories, Pasadena, CA 91101, USA}
\altaffiltext{$\star$}{This paper includes data gathered with the 6.5 meter Magellan Telescopes located at Las Campanas Observatory, Chile.}

\begin{abstract}
There is ongoing debate regarding the extent that environment affects galaxy size growth beyond $z\geq1$.
To investigate the differences in star-forming and quiescent galaxy properties as a function of environment at $z=2.1$, we create a mass-complete sample of 59 cluster galaxies \citep{Spitler2012} and 478 field galaxies with {\M}$\geq9$ using photometric redshifts from the ZFOURGE survey. 
We compare the mass-size relation of field and cluster galaxies using measured galaxy semi-major axis half-light radii ($r_{1/2,maj}$) from CANDELS {\HST}/F160W imaging.
We find consistent mass normalized ({\M}$=10.7$) sizes for quiescent field galaxies ($r_{1/2,maj}=1.81\pm{0.29}${\kpc}) and quiescent cluster galaxies ($r_{1/2,maj}=2.17\pm{0.63}${\kpc}).
The mass normalized size of star-forming cluster galaxies ($r_{1/2,maj}=4.00\pm{0.26}${\kpc}) is $12\%$ larger (KS test $2.1\sigma$) than star-forming field galaxies ($r_{1/2,maj}=3.57\pm{0.10}${\kpc}).  
From the mass-color relation we find that quiescent field galaxies with $9.7\textless${\M}$\leq10.4$ are slightly redder (KS test $3.6\sigma$) than quiescent cluster galaxies, while cluster and field quiescent galaxies with {\M}$\textgreater10.4$ have consistent colors.
We find that star-forming cluster galaxies are on average $20\%$ redder than star-forming field galaxies at all masses.
Furthermore, we stack galaxy images to measure average radial color profiles as a function of mass.
Negative color gradients are only present for massive star-forming field and cluster galaxies with {\M}$\textgreater10.4$, the remaining galaxy masses and types have flat profiles. 
{\ha} given the observed differences in size and color of star-forming field and cluster galaxies, that the environment has begun to influence/accelerate their evolution. 
However, the lack of differences between field and cluster quiescent galaxies indicates that the environment has not begun to significantly influence their evolution at $z\sim2$.
\end{abstract}


\section{Introduction}
Both star-forming and quiescent galaxies exhibit size growth as a function of redshift \citep[e.g.,][]{2005ApJ...626..680D,2010ApJ...713..738W,vanDokkum2010,2013ApJ...766...15P}, with quiescent galaxies growing more rapidly than star-forming galaxies \citep[e.g.,][]{2007MNRAS.382..109T,2013MNRAS.428.1088M,2014ApJ...788...28V}.
The size growth of star-forming galaxies is thought to be fuelled by the addition of new gas that produces stars \citep[e.g.,][]{2013Sci...341...50B} or by minor mergers \citep[e.g.,][]{vanDokkum2010}.
Quiescent galaxies likely grow in size via adiabatic expansion \citep[e.g.,][]{2008ApJ...689L.101F,2010ApJ...718.1460F} or via minor and major mergers \citep[e.g.,][]{2006ApJ...648L..21K,2009ApJ...699L.178N,Guo2011,2013ApJ...763...73S}.
However, it has been suggested that environment may affect or accelerate these growth mechanisms \citep[e.g.,][]{2011MNRAS.415.2993H, Cooper2012,2012MNRAS.419..669M, Raichoor2012, Papovich2012, 2013MNRAS.435..207L, 2013arXiv1310.6754N}.

At $z\textless1$ the density-morphology relation demonstrates that environment can directly affect the properties of galaxies \citep[e.g.,][]{1997ApJ...490..577D,2005ApJ...623..721P,2007ApJ...670..206V}.
This empirical relation shows that gas rich star-forming galaxies are found preferentially in low density environments while gas poor quiescent galaxies are found in the highest density environments.
The models of \citet{2008MNRAS.384....2G} show that mergers and star formation contribute almost equally to the growth rate for galaxies in groups. 
Simulations have shown that the growth of quiescent galaxies could be accelerated in higher density environments where interactions are more frequent and that the most massive galaxies typically reside in the highest galaxy over-densities \citep{2007ApJ...654...53M,Shankar2013}.
The direct comparison of the sizes and light profiles of field and cluster galaxies at $z\textgreater1$ could provide key insight into the epoch and mechanisms where environment begins to affect the size evolution of star-forming and quiescent galaxies.

A handful of studies have shown that the sizes of massive quiescent cluster galaxies are up to $50\%$ larger than quiescent field galaxies at $z\textgreater1$ \citep{Papovich2012,2013ApJ...770...58B,2013arXiv1310.6754N,2014MNRAS.441..203D}.
\citet{2014MNRAS.441..203D} studied the mass-size relation for passive early-type galaxies in clusters at $1.1\textless{z}\textless1.6$ with {\M}$\textgreater10.5$.
From their best-fits of the mass-size relation, they found that median sizes for quiescent cluster galaxies are $30\%$ larger than quiescent field galaxies at the same redshift and mass.
\citet{Papovich2012} and \citet{2013ApJ...770...58B} examined quiescent galaxies in a $z\sim1.6$ cluster and found that cluster galaxies have sizes that are $40\%$ larger than coeval field galaxies.
They attribute some of this difference to recently quenched galaxies on the cluster outskirts with larger effective radii.   
Similarly, \citet{2013MNRAS.435..207L} found that quiescent cluster galaxies at $1\textless{z}\textless2$, with {\M}$\geq11.3$, have effective radii up to $50\%$ larger than field galaxies with similar mass. 
\citet{2013arXiv1310.6754N} compared the sizes of field and cluster early-type galaxies at $z\sim1.8$ and found that the most massive cluster galaxies with {\M}$\textgreater11.5$ were larger than their field counter-parts.
However, \citet{Raichoor2012} measured the mass-size relation for field and cluster quiescent galaxies in a $z=1.3$ cluster and tentatively found that the cluster galaxies have average sizes that are $30\%$ smaller than field galaxies with the same mass and redshift.
Clearly, there is a need for more studies to constrain galaxy size evolution in over-dense regions at $z\textgreater1$.

The study of the size growth of quiescent galaxies as a function of redshift is relevant because they represent a significant fraction of $z=0$ cluster galaxies; however, the majority of high redshift cluster galaxies are star-forming. 
In two spectroscopically confirmed $z\sim2$ galaxy clusters the ratio of star-forming to quiescent galaxies is at least 3:1\citep{2013ApJ...776....9G,2014ApJ...795L..20Y}.
Therefore, analyzing the sizes and light profiles of star-forming cluster galaxies may also be important in distinguishing the effects of environment on galaxy growth.

The majority of environmental and size evolution studies at $z\textgreater1$ quantify the structural properties of individual galaxies using {\HST} \citep{Papovich2012,Raichoor2012,2012ApJ...749..121S,Whitaker2012,2013ApJ...766...15P,2013ApJ...770...58B}.
Due to the low surface brightness and small angular sizes of galaxies at high redshifts, measuring structural parameters for these galaxies is difficult, however, the use of {\HST} imaging provides high resolution and the capability to discern the structural properties of these galaxies.

Image stacking, that represents an average galaxy, can also be used for low mass galaxies to create a deeper image.
Average galaxy radial light profiles measured from image stacks can extend to larger radii.
\citet{vanDokkum2010} used image stacking to study the light profiles of a sample of galaxies with $0\textless{z}\textless2$ to  a surface brightness depth of $\sim28.5$ AB mag arcsec$^{-1}$.
After integrating these light profiles to obtain the radial surface density, they were able to trace the mass growth of galaxies with $11\leq$~{\M}~$\leq11.7$ and found that mass was being gradually added at $r\textgreater5$~kpc.
Mass excess at large radii supports the idea that at least very massive galaxies are growing via minor mergers from $z\sim2$ \citep{2010ApJ...715..202H,2013MNRAS.429.2924H,Shankar2013}.
By comparing the sizes and light profiles of field and cluster star-forming and quiescent galaxies at high redshift, a more comprehensive picture of size growth and the underlying mechanisms may be gained.

A large sample of both field and cluster star-forming and quiescent galaxies is necessary to constrain the epoch at which environment began to affect galaxy growth; however, acquiring large samples of field and cluster galaxies at high redshift is difficult.
First, clusters become increasingly rare at $z\textgreater1$.
Secondly, current broad-band photometric samples encounter larger redshift errors and are unable to successfully identify galaxy over-densities.
However, the FourStar Galaxy Evolution (ZFOURGE) survey has produced accurate photometric redshifts which allow for environment to be determined as well as other galaxy properties derived from photometric fits (Straatman et al. 2015, in prep.), as was shown from the discovery of a $z=2.1$ galaxy cluster \citep{Spitler2012,2014ApJ...795L..20Y}.

In this paper, we study for the first time, the mass-size relation for star-forming and quiescent field and cluster galaxies at $z=2.1$ obtained from the ZFOURGE survey with {\M}$\geq9$.
We cross-matched the ZFOURGE catalog with the size and {\n} index measurements of \citet{2014ApJ...788...28V}, based on the 3DHST survey \citep{2014arXiv1403.3689S}.
The mass-color relation for star-forming and quiescent field and cluster galaxies is also examined. 
In addition, we measure stacked radial color profiles of our sample of galaxies using the {\HST}/WFC3 F160W and {\HST}/ACS F814W images.
We compare the individual colors, radial color profiles, sizes, and {\n} indices of average field and cluster galaxies to determine the effects of environment on star-forming and quiescent galaxies.
The paper is organized as follows: in Section \ref{sec:sample} we describe our sample selection and its properties, in Section \ref{sec:anal} we describe our construction of the mass-size relation, mass-color relation and radial color profiles, results are discussed in Section \ref{sec:dis}, and we present our concluding remarks in Section \ref{sec:con}.

Throughout our study we assume a $\Lambda$CDM cosmology with $\Omega_{\Lambda}=0.73$, $\Omega_{m}=0.27$, and $H_{0}=71$ km s$^{-1}$.

\section{The Sample}
\label{sec:sample}
\subsection{ZFOURGE Imaging and Catalog}
The ground-based near-infrared imaging data was taken as part of the FourStar Galaxy Evolution survey (ZFOURGE; Straatman et al. 2015, in prep.) during 2011-2012, using the Fourstar instrument \citep{2013PASP..125..654P} on the 6.5~m Magellan telescope at Las Campanas Observatory in Chile.
As part of this survey, $11' ~\times~11'$ areas in the COSMOS \citep{2007ASPC..375..166S}, UDS \citep{2007MNRAS.379.1599L}, and CDFS \citep{2002ApJS..139..369G} fields were targeted using medium-bands J$_{1}$, J$_{2}$, J$_{3}$, H$_{s}$, H$_{l}$, and the K$_{s}$ broad-band.
The combined length of observations was $\approx$ 70 hours which translates to 5$\sigma$ depths of $\sim$ 25.5 AB mag in J$_{1}$, J$_{2}$, J$_{3}$ and $\sim$ 25 AB mag in H$_{s}$, H$_{l}$, and K$_{s}$ \citep{2013ApJ...768...56T}.
\citet{2014arXiv1412.3806P} found that the ZFOURGE data are $80\%$ complete at Ks(AB) = 24.5, 24.7, and 25.1 for the CDF-S, COSMOS, and UDS fields.

The raw imaging data was processed with a modified pipeline based on that of the NEWFIRM survey \citep{2011ApJ...735...86W}. 
The reduction process for the data will be fully detailed in Straatman et al. (in prep.). 
The point-spread-function (PSF) FWHM of the ZFOURGE K$_{s}$-band image is  $0.4''$. 

Along with the medium-bands and the K$_{s}$-band, multi-wavelength data covering COSMOS was used when preforming SED fits. 
In total, there were 34 photometric bands spanning rest-frame wavelengths of $\sim 0.1-2.7\mu$m. 
Photometric redshifts and rest-frame colors were measured by fitting template SEDs to PSF-matched optical-NIR photometry with the SED-fitting code {\tt EAZY} \citep{Brammer2008}.
Stellar masses were obtained by using {\tt FAST} \citep{Kriek2009} to fit stellar population synthesis templates to the same photometry.
Stellar population models were made with the population synthesis code of Bruzual \& Charlot (2003) assuming a Chabrier IMF and solar metallicity.
Star formation histories were modelled as exponentially decreasing ($\Psi~{\propto}~e^{-t/\tau}$) with values of log($\tau$/year)=7-11 in steps of 0.2 and log(age/yr)$=7.5-10.1$ insteps of 0.1. 
The derived photometric redshift uncertainties of the ZFOURGE are $\delta$$z/(1+z)\textless0.02$ \citep{2014ApJ...792..103K,2014ApJ...783...85T,2014ApJ...795L..20Y}.
The combined redshift and mass uncertainty ranges from $5-15\%$ over the redshift range of 0.5-3 \citep{2014ApJ...783...85T}. 
\begin{figure*}[]
\epsscale{1.2}\plotone{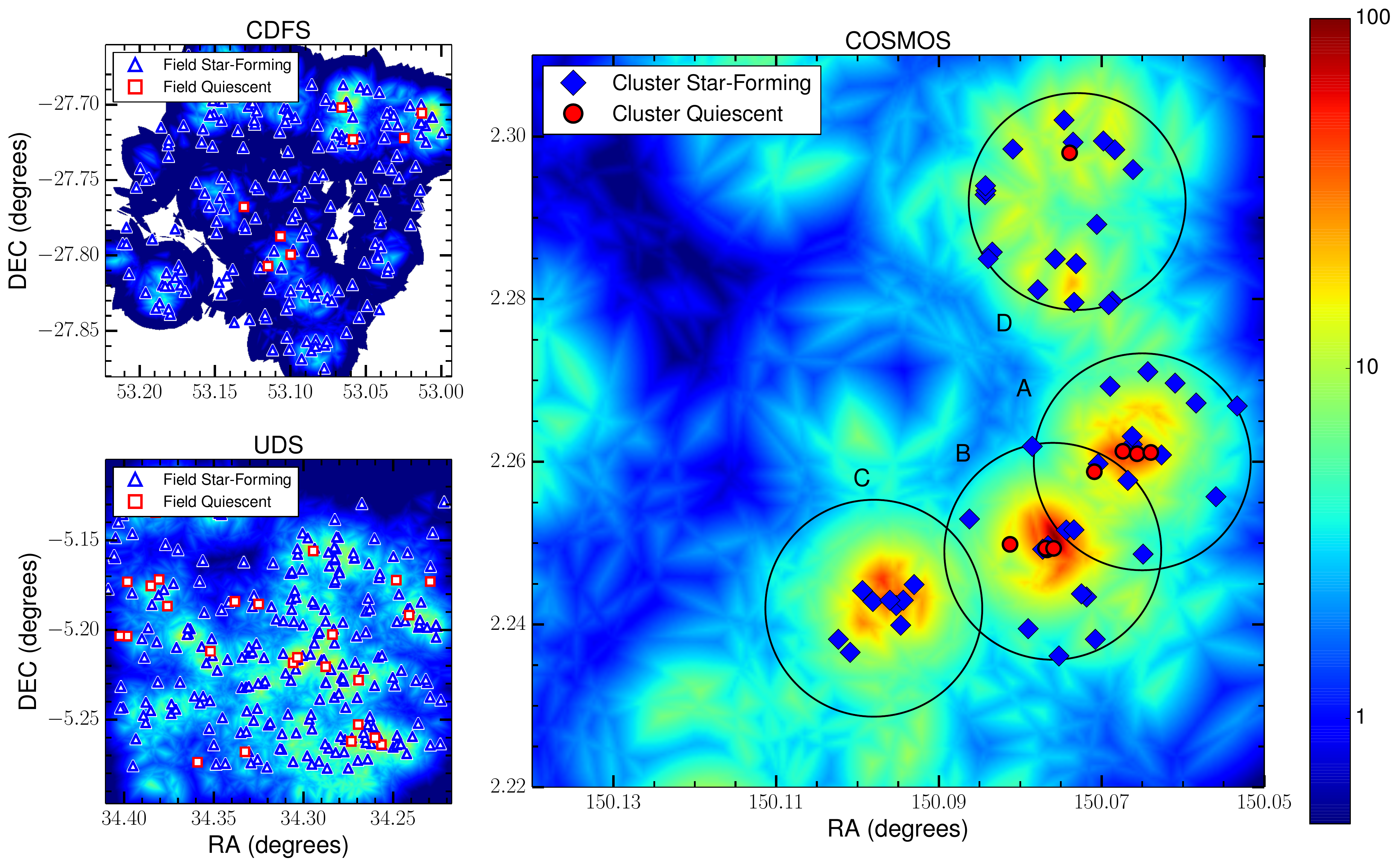}
  \caption{Seventh-nearest-neighbor projected density maps of CDFS, UDS, and COSMOS fields. The color bar represents the significance, in sigma, of the projected density of CDFS, UDS, or COSMOS at $2.0\leq{z}\leq2.2$ above the mean density. The mean density is averaged over all three fields at $1.8\leq{z}\leq2.0$ and $2.2\leq{z}\leq2.4$. Field star-forming galaxies (open blue triangles) and field quiescent galaxies (open red squares) were selected from the CDFS and UDS fields, where no significant ($\textgreater10\sigma$) large-scale over-densities were found. Right: Four galaxy over-densities in the COSMOS field were found, using the seventh-nearest-neighbor metric, with photometric redshifts between $2.0\leq{z}\leq2.2$. Regions A, B, and C are the original over-densities identified in \citet{Spitler2012}. Region D was identified post-publication. Our sample of cluster galaxies is shown as blue diamonds (star-forming galaxies) and red circles (quiescent galaxies) within $48''$($\sim$ 400 kpc at $z=2.1$) apertures (black circles).}
  \label{fig:dens}
\end{figure*}

\subsection{Galaxy Structural Parameters}
\label{sec:struc}
\citet{2014ApJ...788...28V} used F125W, F140W, and F160W {\HST} imaging from CANDELS \citep{2011ApJS..197...36K,2011ApJS..197...35G} as well as the 3D-HST catalog \citep{2014arXiv1403.3689S} to create catalogs of individual galaxy structural parameters.
They used {\tt GALAPAGOS}, which incorporates both {\tt SExtractor} \citep{1996A&AS..117..393B}  and {\tt GALFIT} \citep{2010ApJ...721..193P}, to detect and model galaxies. 
The galaxies were fit using a single {\n} fit, with custom made PSFs for each field, and a limited range of best-fit values for structural parameters such as half-light radius ($0.3-400$ pixels), {\n} index ($0.2-8$), and axis ratio ($0.0001-1$).
Measurement uncertainties were derived by rerunning {\tt GALAPAGOS} on the same object over varying image depths; the full details of the parameter fitting can be found in \citet{2012ApJS..203...24V}.
Reliable fits are flagged with f=0 or 1 and unreliable fits are flagged f$\geq2$.
Unreliable fits are flagged typically due to low SNR and blending of objects and we exclude these galaxies from size analysis.
They determined that reliable (accuracy $\leq20\%$) half-light radii, $r_{e}$, and {\n} indices, n, can be derived for galaxies with F160W magnitudes of 24.5 and 23.5 AB magnitudes or brighter, respectively. 
We discuss the F160W magnitude distribution of our sample near the end of this section.

These catalogs include sizes for galaxies within COSMOS, UDS, and CDFS and are the largest and most accurate catalogs to date for which we can study the size evolution of galaxies.
Therefore, we cross-matched the ZFOURGE galaxy sample with the van der Wel F160W catalog for the three fields to create a value added catalog that includes sizes for 75 $\%$ of the ZFOURGE catalogs.
We were not able to match 100$\%$ of the ZFOURGE galaxies because the ZFOURGE survey footprint is slightly different than the CANDELS image footprints used to create the \citet{2014ApJ...788...28V} galaxy catalogs.

\subsection{Field \& Cluster Galaxy Selection}
The z=2.095 cluster has 57 spectroscopically confirmed members obtained by \citet{2014ApJ...795L..20Y} using MOSFIRE on Keck I.
The cluster has a velocity dispersion of 552 km/s and is likely a Virgo-like cluster progenitor.
However, given the spectroscopic bias toward strong emission-line galaxies, and that \citet{2014ApJ...795L..20Y} confirm that the photometric redshifts of ZFOURGE are accurate to within $2\%$, we choose to use photometric redshifts for our sample selection.
This will provide a more uniform selection of all galaxy types.
The spectroscopic redshift and photometric redshift cluster contamination is discussed at the end of this section.

We constructed a mass complete sample of cluster galaxies in the redshift range of the cluster found in the ZFOURGE survey at $2\leq{z}\leq2.2$ \citep{Spitler2012} with {\M}~$\geq9$ (for m$_{K}\leq24.5$ AB mag) to examine the effects of environment on the evolution of galaxies.  
This cluster consists of three over-densities, A, B, and C that were originally identified using the seventh-nearest neighbour metric \citep[see][for details]{Spitler2012}.
Post-publication, and after catalog refinements, an additional over-density, D, was discovered at the same redshift range and in the vicinity as the three listed in \citet{Spitler2012} and is included in our study. 
In Figure \ref{fig:dens} we show a revised seventh-nearest-neighbor projected density map of the significance, in sigma, of COSMOS at $2.0\leq{z}\leq2.2$ above the mean density. 
The mean density is averaged over all three fields at $1.8\leq{z}\leq2.0$ and $2.2\leq{z}\leq2.4$.

\begin{figure}[]
\centering
\epsscale{1.15}\plotone{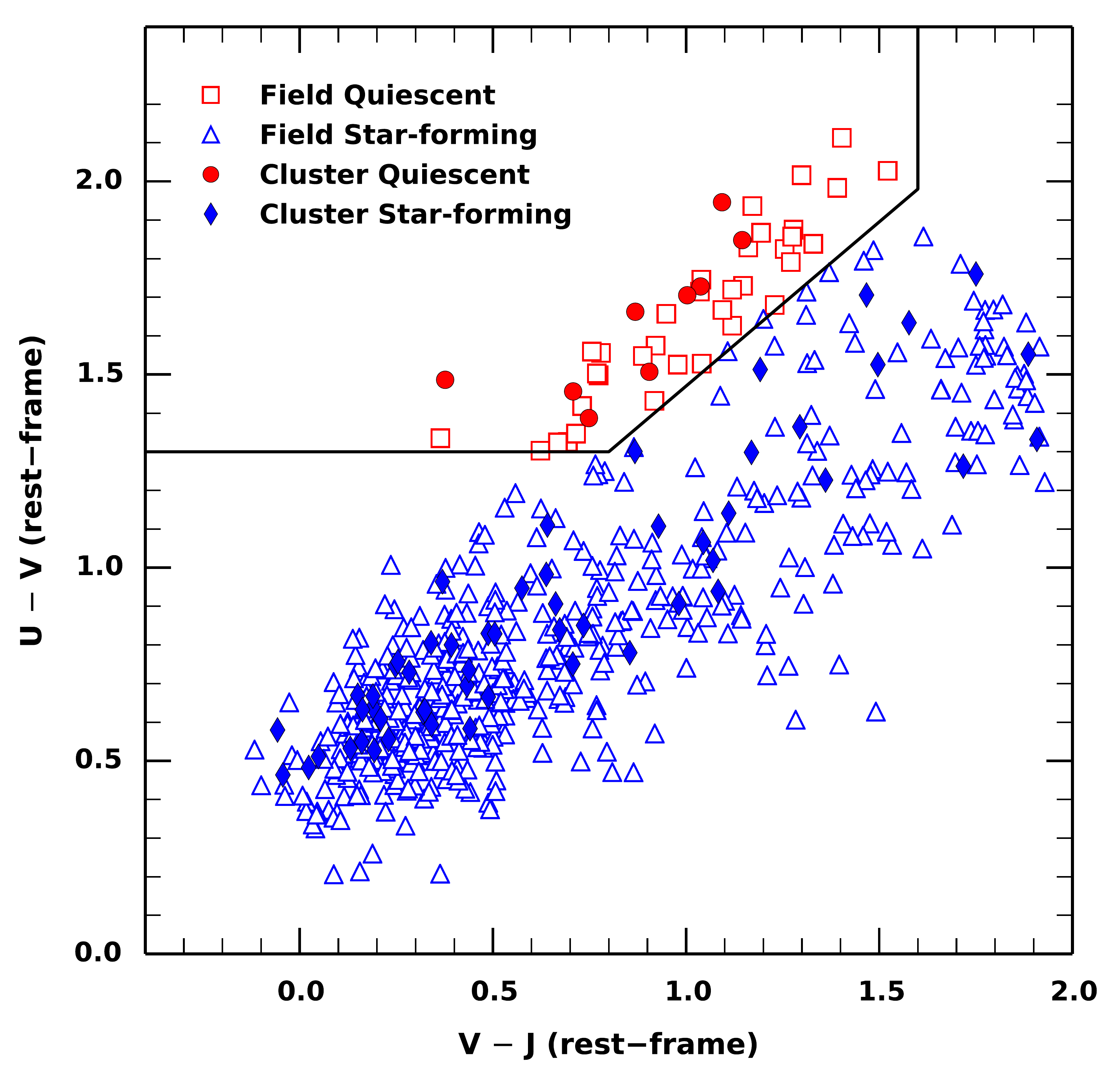}
   \caption{Rest-frame U~$-$~V versus V~$-$~J colors for our sample of field and cluster galaxies at $z=2.1$. Star-forming cluster (field) galaxies are shown as filled (open) blue diamonds (triangles). Quiescent cluster (field) galaxies are shown as filled (open) red circles (squares). The black line represents the boundary for quiescent galaxies (above) and star-forming galaxies (below) as defined by \citet{Spitler2012}.}
      \label{fig:UVJ}
\end{figure}
\begin{figure}[]
\centering
\epsscale{1.2}\plotone{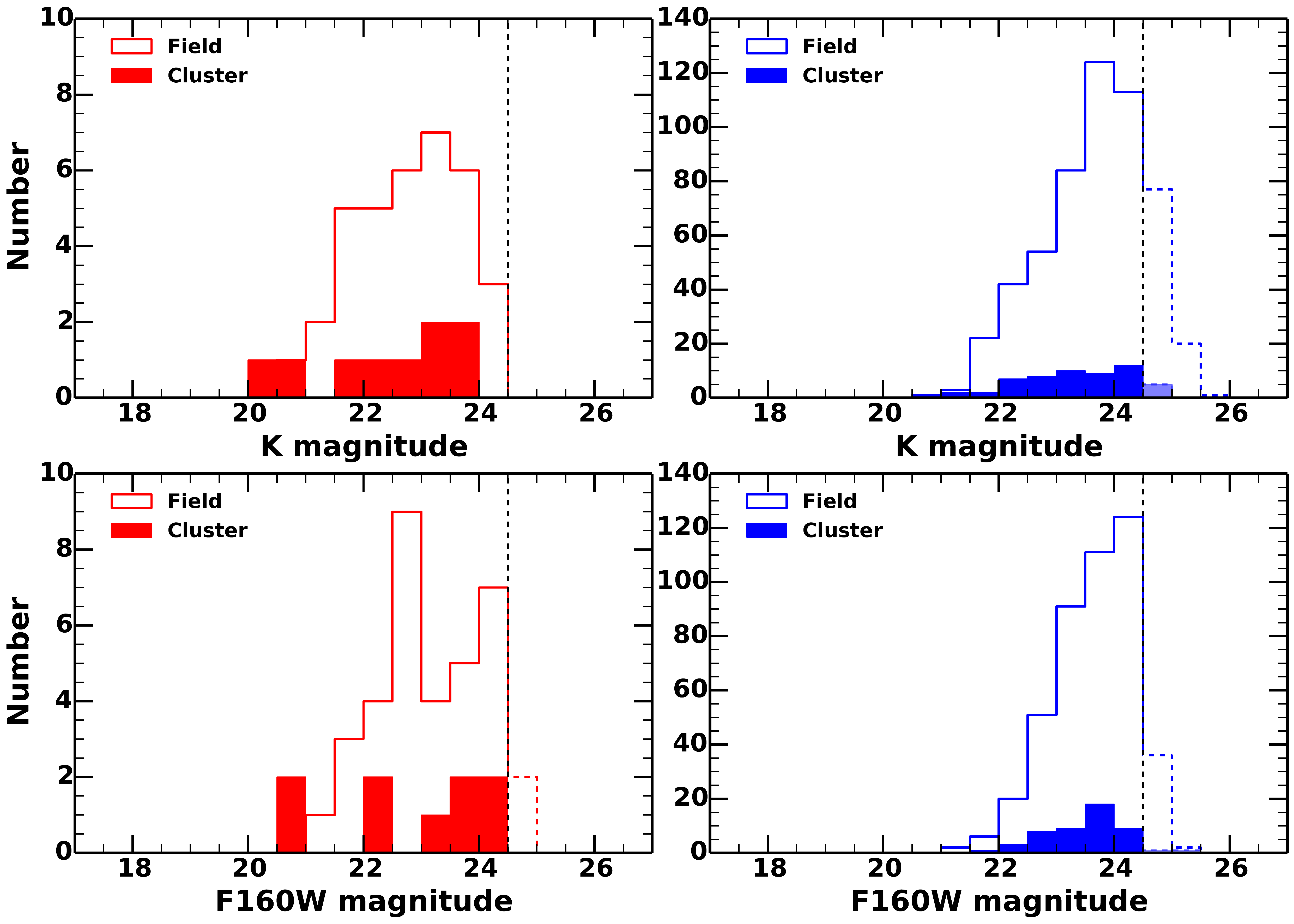}
   \caption{Top: Distributions of m$_{K}$ for our sample of field (open histograms) and cluster (closed histograms) quiescent (left panel) and star-forming (right panel) galaxies. The dashed and light color histograms show where the field and cluster samples fall below the K-band magnitude limit of m$_{K}\textgreater24.5$ AB mag (for {\M}$\geq9$) shown as a dashed black line. We remove galaxies that fall below this limit. Bottom: Distributions of m$_{F160W}$ magnitudes for our sample of field (open histograms) and cluster (closed histograms) quiescent (left panel) and star-forming (right panel) galaxies. The dashed and light color histograms show where the field and cluster samples fall below the m$_{F160W}$ magnitude limit for reliable sizes, shown as a dashed black line.}
      \label{fig:mags}
\end{figure}
The over-density has a complex structure, and in order to maximize the amount of structure included in our cluster galaxy selection we define the center of D to be RA~$=$~10:00:17.520, DEC~$=$~$+$02:17:31.20 (J2000).
Using these center coordinates as well as the centroids for the original over-densities from \citet{Spitler2012}, cluster members were defined as galaxies within $48''$($\sim$ 400 kpc) of these coordinates and within the redshift range $2\leq{z}\leq2.2$. 
In Figure \ref{fig:dens}, we show our sample of cluster galaxies as filled red circles (quiescent galaxies) and filled blue diamonds (star-forming galaxies) within $48''$ apertures (black circles). 
We selected $48''$ apertures for the cluster distances so as to minimize contamination of field galaxies and maximize the amount of cluster members. 

We selected our field galaxy sample from the UDS and CDFS fields because of the known over-density at $2\leq{z}\leq2.2$ in the COSMOS field.
Field galaxies were selected using the same redshift and mass limits as the cluster galaxy sample.
In Figure \ref{fig:dens} we show a revised seventh-nearest-neighbor projected density map of the significance, in sigma, of CDFS ad UDS at $2.0\leq{z}\leq2.2$ above the mean density. 
The mean density is averaged over all three fields at $1.8\leq{z}\leq2.0$ and $2.2\leq{z}\leq2.4$.
We use this to confirm no significant ($\textgreater10\sigma$) large-scale over-densities exist in UDS and CDFS at $2\leq{z}\leq2.2$.

The cluster contamination fraction was estimated by two different methods.
First, we calculated the number density of field galaxies in UDS and CDFS in the redshift range of the cluster in COSMOS and then divided this by the cluster number density. 
We then determine a cluster contamination fraction of 0.25 for star-forming galaxies and 0.1 for quiescent galaxies.
The second method for estimating the cluster contamination came from using the high confidence spectroscopic redshifts of \citet{2014ApJ...795L..20Y}.
Out of the total photometric cluster sample, 64, 16 galaxies have spectroscopic redshifts not within the cluster.
Therefore, we can estimate the contamination is 16/64 or $25\%$ which is equivalent to the contamination estimated for the star-forming photometric sample.
\begin{figure*}[ht]
\epsscale{1.17}\plotone{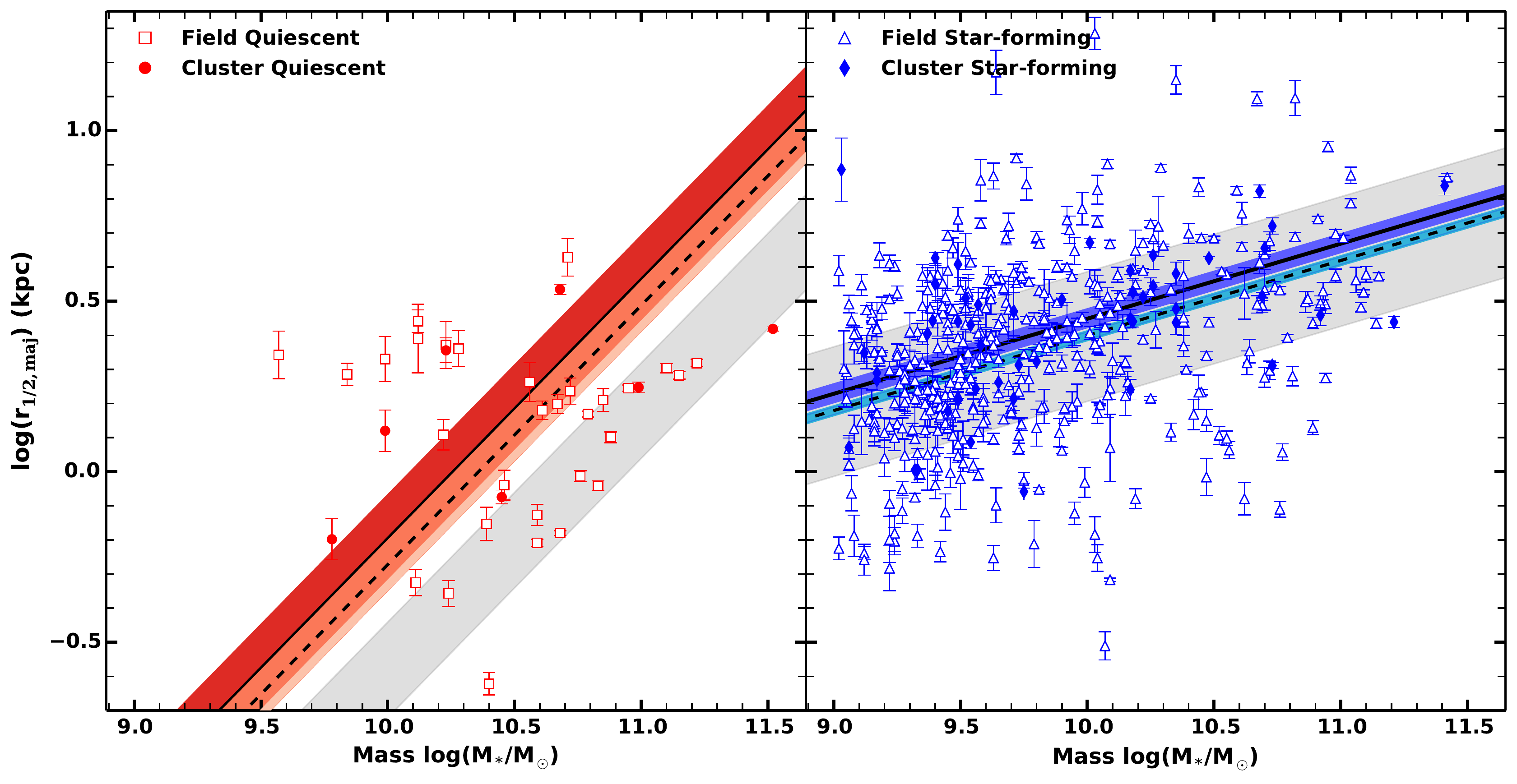}
\caption{Mass-size relation for our sample of quiescent (left) and star-forming (right) field and cluster galaxies at $z=2.1$. Star-forming field (cluster) galaxies are shown as open (filled) blue triangles (diamonds). Quiescent field (cluster) galaxies are shown as open (filled) red squares (circles). Our fits are calculated using the parametrized-fit slope of \citet{2014ApJ...788...28V} for star-forming and quiescent galaxies at $z=2.25$. The fits are shown as dashed (solid) lines for field (cluster) galaxies. The errors were calculated from bootstrapping the fit and the distribution of fits was non-gaussian, therefore we use percentiles equivalent to $1\sigma$ for the errors, shown as light (dark) contours for field (cluster) galaxies. We show the best-fits for star-forming and quiescent galaxies at $z=2.25$ from \citet{2014ApJ...788...28V} as grey contours which include the $1\sigma$ scatter. There is a discrepancy between the fit to our data and the best-fit of \citet{2014ApJ...788...28V} for quiescent galaxies due to their exclusion of galaxies with {\M}$\textless10.3$. If we exclude quiescent galaxies with {\M}$\textless10.3$ from our fit, our sizes are consistent with \citet{2014ApJ...788...28V} (see text).} Our fits indicate that quiescent cluster and quiescent field galaxies are consistent in size. Star-forming cluster galaxies are larger in size than field star-forming galaxies by $2.4\sigma$.
  \label{fig:svm}
\end{figure*}

The U~$-$~V versus V~$-$~J rest-frame color-color diagram has been shown to efficiently separate quiescent galaxies from star-forming galaxies when accurate rest-frame colors are used \citep[e.g.,][]{Williams2009,2009ApJ...706..885W,Whitaker2012,2014MNRAS.440.1880W}. 
Specifically, older stellar populations with strong Balmer breaks (4000\AA) are characterised by red U~$-$~V colors and blue V~$-$~J colors. 
From {\tt EAZY} SED fits of the photometric data, we were able to calculate rest-frame colors for the galaxies in our sample.
We separate our field and cluster galaxies into star-forming and quiescent using the quiescent selection box on the UVJ color relation (defined by (U~$-$~V)~$\textgreater0.87~\times$~(V~$-$~J)~$+~0.60$, (U~$-$~V)~$\textgreater1.3$, and (V~$-$~J)~$\textless1.6$).
Galaxies that lie above this diagonal are classified as quiescent.
The UVJ relation for our sample is shown in Figure~\ref{fig:UVJ}.
The cluster sample contains 9 quiescent galaxies and 55 star forming galaxies.
The field sample contains 35 quiescent galaxies and 541 star forming galaxies.

At $z\sim2$ ZFOURGE is mass complete to masses of {\M}$=9$ for m$_{K}\leq24.5$ AB mag (Straatman et al. 2015, in prep).
\citet{2014arXiv1412.3806P} find that the ZFOURGE data are at least $80\%$ complete at this depth in all three fields.
In Figure \ref{fig:mags} we show the distribution of m$_{K}$ for field and cluster star-forming and quiescent galaxies.
We find that ~100 of the star-forming field galaxies and 5 star-forming cluster galaxies fall below the ZFOURGE m$_{K}$ selection magnitude limit and were removed from the sample.
All of the quiescent galaxies are above the magnitude limit.
In the lower panels of Figure~\ref{fig:mags} we show the distribution of m$_{F160W}$ magnitudes of our sample. 
If we remove galaxies that are fainter than m$_{K}=24.5$ AB mag from our sample we have 40 star-forming and 2 quiescent field galaxies that lie below the F160w magnitude limit for reliable sizes, see Figure~\ref{fig:mags} bottom panel.
These galaxies will have larger uncertainties in their sizes, however, we do not remove them from the sample because we weight by error in size when calculating our median sizes.
The final sample size for each environment and galaxy type is show in Table 1.
\begin{table*}[]
\begin{center}
\caption{{\n} indices and mass-normalised median sizes of Star-forming and Quiescent \\ Field and Cluster Galaxies derived from {\HST}/WFC3 F160W images}
\label{table:gal}
	\begin{tabular}{c c c c c c c}
	\hline\hline
	&Quiescent&&&Star-forming  \\[0.5ex]
	\hline
	Environment & Fraction* & $\phantom{-}$$r_{1/2, maj}$ $\phantom{(0.40\sigma)}$& $\phantom{-}n$ $\phantom{(0.14\sigma)}$& Fraction &  $\phantom{-}$$r_{1/2, maj}$ $\phantom{(2.4\sigma)}$& $n$ $\phantom{(1\sigma)}$\\[0.5ex]
	&&$\phantom{-}$(kpc) $\phantom{(0.40\sigma)}$&&& $\phantom{-0}$(kpc) $\phantom{(0.40\sigma)}$&\\[0.25ex]
	\hline\\[-1.5ex]
	Field & 30/35 & $\phantom{-}1.81\pm{0.29}$ $\phantom{(0.40\sigma)}$& $\phantom{-}3.39\pm{0.34}$ $\phantom{(0.14\sigma)}$& 410/443 & $3.57\pm{0.10}$ $\phantom{(2.4\sigma)}$& $1.64\pm{0.07}$ $\phantom{(1\sigma)}$\\[0.5ex]
	Cluster & 7/9 & $\phantom{-}2.17\pm{0.63}$ $\phantom{(0.40\sigma)}$& $\phantom{-}3.49\pm{0.66}$ $\phantom{(0.14\sigma)}$& 49/50 & $\phantom{-}4.00\pm{0.26}$ $\phantom{(1.54\sigma)}$& $1.47\pm{0.19}$ $\phantom{(1\sigma)}$\\[0.5ex]
	\\[-2.5ex]
	$\Delta_{FC}$$^{**}$ &&$-0.36\pm{0.69}$ ($0.52\sigma$)&$-0.10\pm{0.74}$ ($0.14\sigma$)&& $-0.43\pm{0.28}$ ($1.54\sigma$)&$\phantom{0.0}$$0.17~\pm~{0.20}$~($0.84\sigma$)\\
	\hline \\
	\end{tabular}
	\end{center}
	\vglue -2ex
      		$\phantom{000000000}$$^{*}$ Fraction of objects from \citet{2014ApJ...788...28V} with reliable fits, quality flag = 0,1.\\
			$\phantom{000000000}$$^{**}$ $\Delta_{FC}\equiv$ Field~$-$~Cluster
\end{table*}

\section{Analysis \& Results}
\label{sec:anal}
\subsection{Mass-Size Relation}
In Figure \ref{fig:svm} we show the mass-size distribution of the field and cluster galaxies. 
We define galaxy size as the half-light effective radii measured along the semi-major axis, $r_{1/2, maj}$, obtained from the \citet{2014ApJ...788...28V} size catalog.
The effective radii are measured using {\tt GALFIT} and we only use objects that were flagged to have reliable structural parameters.
The fraction of galaxies used for the median size calculation for each sample is shown in Table \ref{table:gal}.

In order to determine if cluster and field galaxies differ in their sizes, we use the same parameterization as \citet{2014ApJ...788...28V} to fit for the mean size as a function of mass:
\begin{equation} r~(m_{*})/\text{kpc}=A~{\cdot}~\text{m}_{*}^{\alpha} \end{equation}
where $m_{*}~{\equiv}~M_{*}/5\times10^{10}~$M$_{\odot}$ and is the same mass normalization used by \citet{2014ApJ...788...28V}.
We adopt the slope of the mass-size relation, $\alpha$, of van der Wel et al. ($0.76\pm{0.04}$  and $0.22\pm{0.01}$ for star-forming and quiescent galaxies, respectively) and simply fit for the y-intercept, $A$. 
Errors in the mean size are determined from bootstrapping the fit for $A$.
The mass normalised mean sizes, for $m_{*}~{\equiv}~M_{*}/5\times10^{10}~$M$_{\odot}$, derived from the best fits and their errors are shown in Table~\ref {table:gal}. 

In Figure \ref{fig:svm}, we show the best-fits to the sizes of field and cluster galaxies along with bootstrap derived errors. 
The best-fit normalization that we find for the star-forming galaxies is consistent with that found by \citet{2014ApJ...788...28V} for star-forming galaxies at $z=2.25$.
For both field and cluster quiescent galaxies, our best fit is offset to larger sizes relative to the best fit derived by \citet{2014ApJ...788...28V}.
This offset is due to the fact that their fit includes a morphological misclassification fraction and mass limit of {\M}$\textgreater10.3$ which excludes objects that scatter to the upper left region of the size-mass relation.
If we exclude quiescent galaxies with {\M}$\textless10.3$ from our fit, we find field and cluster mass-normalized sizes of $1.13\pm{0.14}${\kpc} and $1.32\substack{+0.52\\ -0.38}${\kpc}, respectively, which are consistent with the sizes of quiescent galaxies at z=2.25 found by \citet{2014ApJ...788...28V}.
Given we are interested in determining the difference between the average sizes of field and cluster galaxies, and assuming that both the field and cluster quiescent galaxy populations with {\M}$\textless10.3$ are represented, we include all galaxies that have colors consistent with quiescent galaxies and with {\M}$\geq9$ in our fit.
The mass-normalised sizes are listed in Table \ref{table:gal}. 

We found that the mean sizes of star-forming cluster galaxies are $12\%$ larger than the mean sizes of field star forming galaxies.
Cluster star-forming galaxies with {\M}$\geq9$ have typical sizes of 4.00$\pm{0.26}$ kpc and field galaxies have typical sizes of 3.57$\pm{0.10}$ kpc.  
A Kolmogorov-Smirnov (KS) test indicates that the star-forming field and cluster size distributions differ by $2.1\sigma$.

The mean sizes of quiescent field and cluster galaxies with {\M}$\geq9$ are consistent within the errors with cluster galaxies having typical sizes of 2.17$\pm{0.63}$ kpc and field galaxies of 1.81$\pm{0.29}$ kpc. Our sample of cluster quiescent galaxies is small, however, we only have one galaxy with {\M}$\sim10.5$ and $r_{1/2, maj}\sim1${\kpc}, suggesting a lack of compact massive cluster quiescent galaxies. 

In Section \ref{sec:dis}, we review the effect that a large error in the median size of the cluster quiescent galaxies has on the sensitivity of detecting a difference in median sizes of quiescent field and cluster galaxies.
\begin{figure}[]
\epsscale{1.15}\plotone{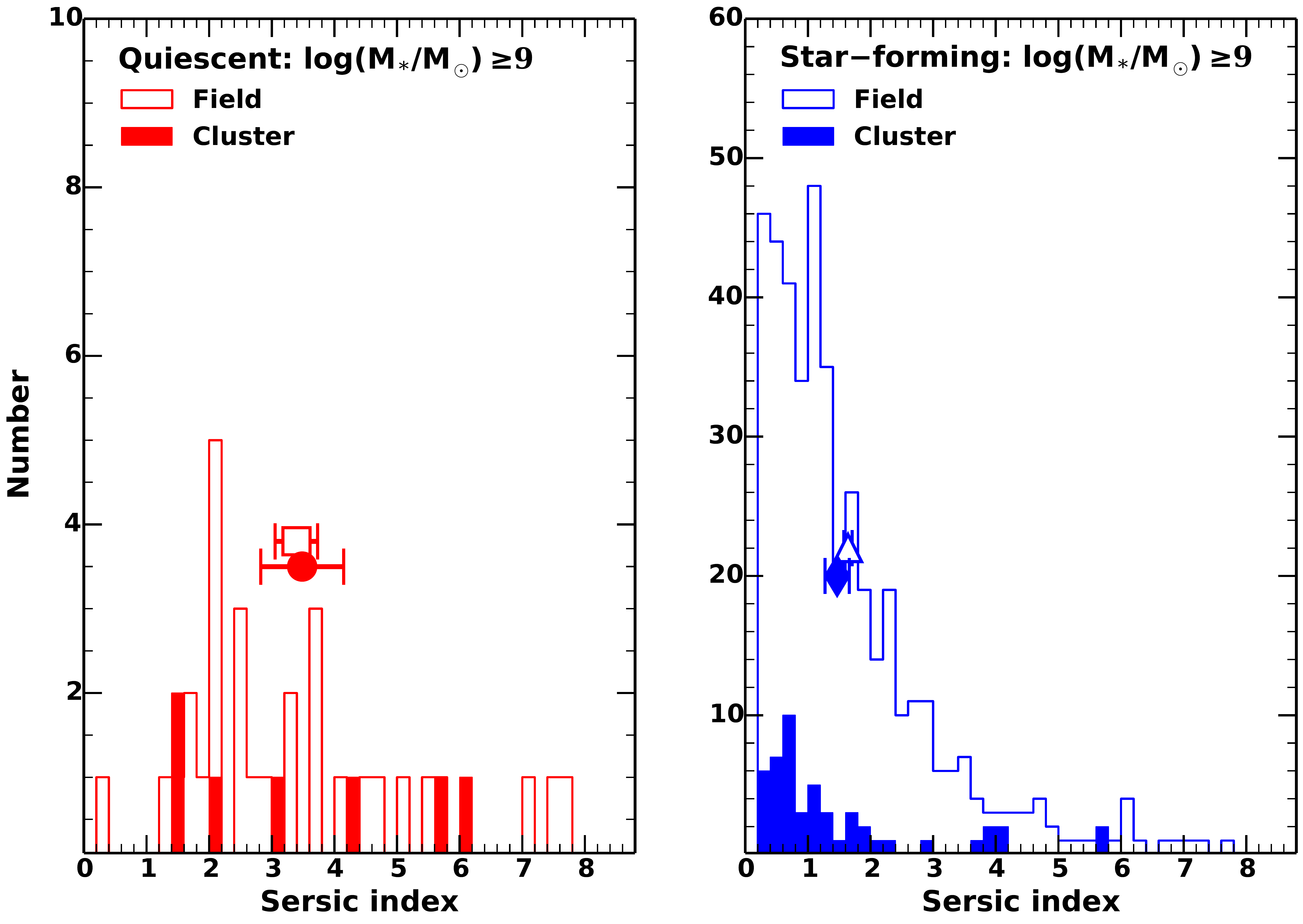}
\caption{Distribution of {\n} indices of field and cluster, star-forming (blue) and quiescent (red) galaxies. The left panel shows cluster galaxies (solid histogram) and the right panel shows field galaxies (open histogram). The average {\n} index with one sigma error for each sample is shown with the same symbol and color coding as Figure \ref{fig:svm} (the field values are offset by 1 in y-space so that they can be distinguished from the cluster average). {\n} indices have a similar distribution for field and cluster galaxies.}
  \label{fig:s}
\end{figure}
\subsection{{\n} indices}
In addition to galaxy size as a function of environment we examine the distribution of {\n} indices for our field and cluster samples shown in Figure \ref{fig:s}.
One constraint used by \citet{2014ApJ...788...28V} during the {\tt GALFIT} fitting process was that {\n} values where fixed to a set range from $n=0.2-8$.
While the majority of single sersic fits have proven to provide reasonable fits for galaxy structural parameters at high redshifts, occasionally some galaxies may be better fit with a double component \citep[e.g.,][]{Raichoor2012}.
These galaxies have $n$ values equal to the boundary at $n=8$. 
Here, we removed these unrealistic $n=8$ galaxies before taking the average {\n} index for each sample.
We use the error in the mean for the error in the average {\n} index.
We found that the average {\n} indices of quiescent field, $n=3.39\pm{0.34}$ and cluster, $n=3.49\pm{0.66}$, galaxies are consistent.
The {\n} indices of star-forming field galaxies, $n=1.64\pm{0.07}$ are consistent with the {\n} indices of star-forming cluster galaxies, $n=1.47\pm{0.19}$.
We note that some of the field quiescent and field and cluster star-forming galaxies have m$_{F160W}$ below 23.5 AB mags, which is the magnitude limit for reliable $\leq20\%$ {\n} indices. 
However, when we remove these galaxies, the distribution and median {\n} do not change.
The median {\n} values and their errors are listed in Table \ref{table:gal}.

\subsection{Colors}
In order to determine if the stellar populations of field and cluster galaxies differ, we examine their individual integrated colors.
We use the CANDELS F814W {\HST}/ACS ($\lambda\sim$0.26 $\mu$m rest-frame) and F160W {\HST}/WFC3 ($\lambda\sim$0.48 $\mu$m rest-frame) images which contain our galaxy sample. 
The two {\HST} images have different point spread functions (PSFs), therefore we used a F814W {\HST}/ACS image PSF matched to F160W {\HST}/WFC3 from 3D-HST \citep{2012ApJS..200...13B}.
To obtain individual galaxy colors we first create image thumbnails of $60\times60$ pixels, or $\sim 30 \times 30$ kpc at $z=2.1$, for each galaxy from both the {\HST} images.
\begin{figure}[]
\epsscale{1.15}\plotone{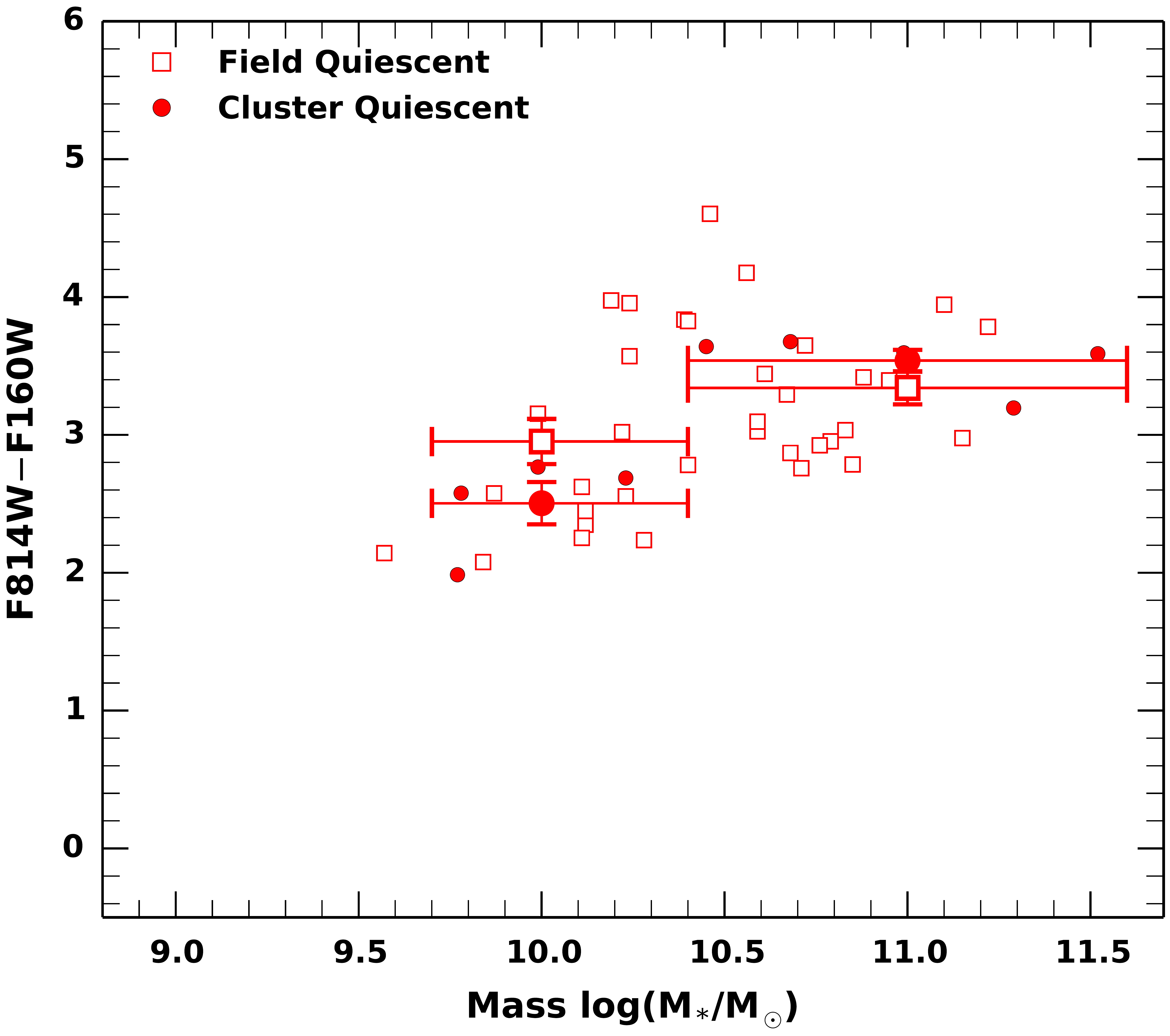}\\
\epsscale{1.15}\plotone{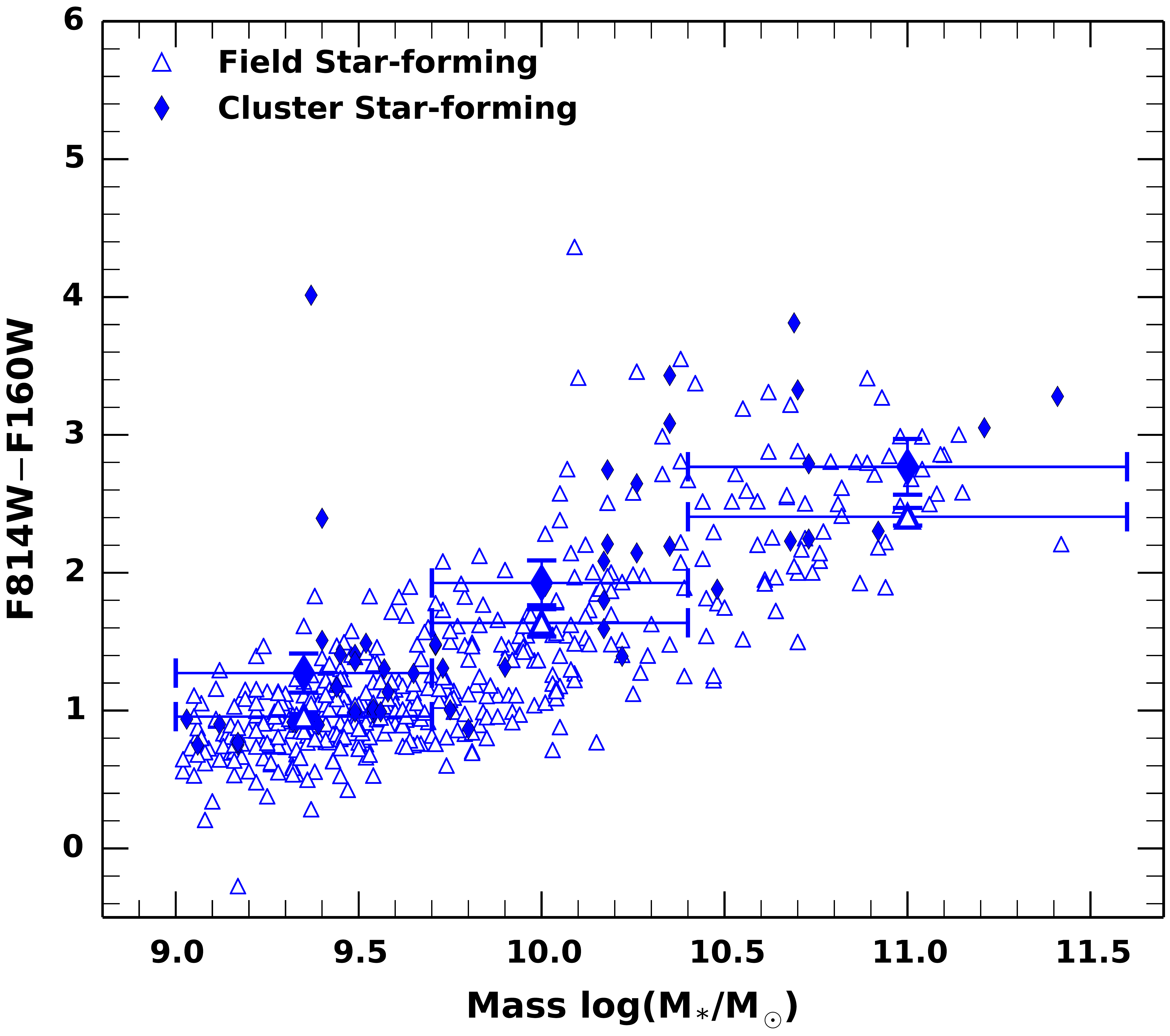}
\caption{F814W$-$F160W observed color versus mass for field and cluster quiescent (left panel) and star-forming (right panel). The large symbols represent the average color for each sample of galaxies separated by mass. Error bars in the x-direction denote the width of the mass bin included in the average. The y-error bars represent the standard deviation in the mean color.}
  \label{fig:mvc}
\end{figure}
\begin{figure*}[]
\centering
  \epsscale{1}\plotone{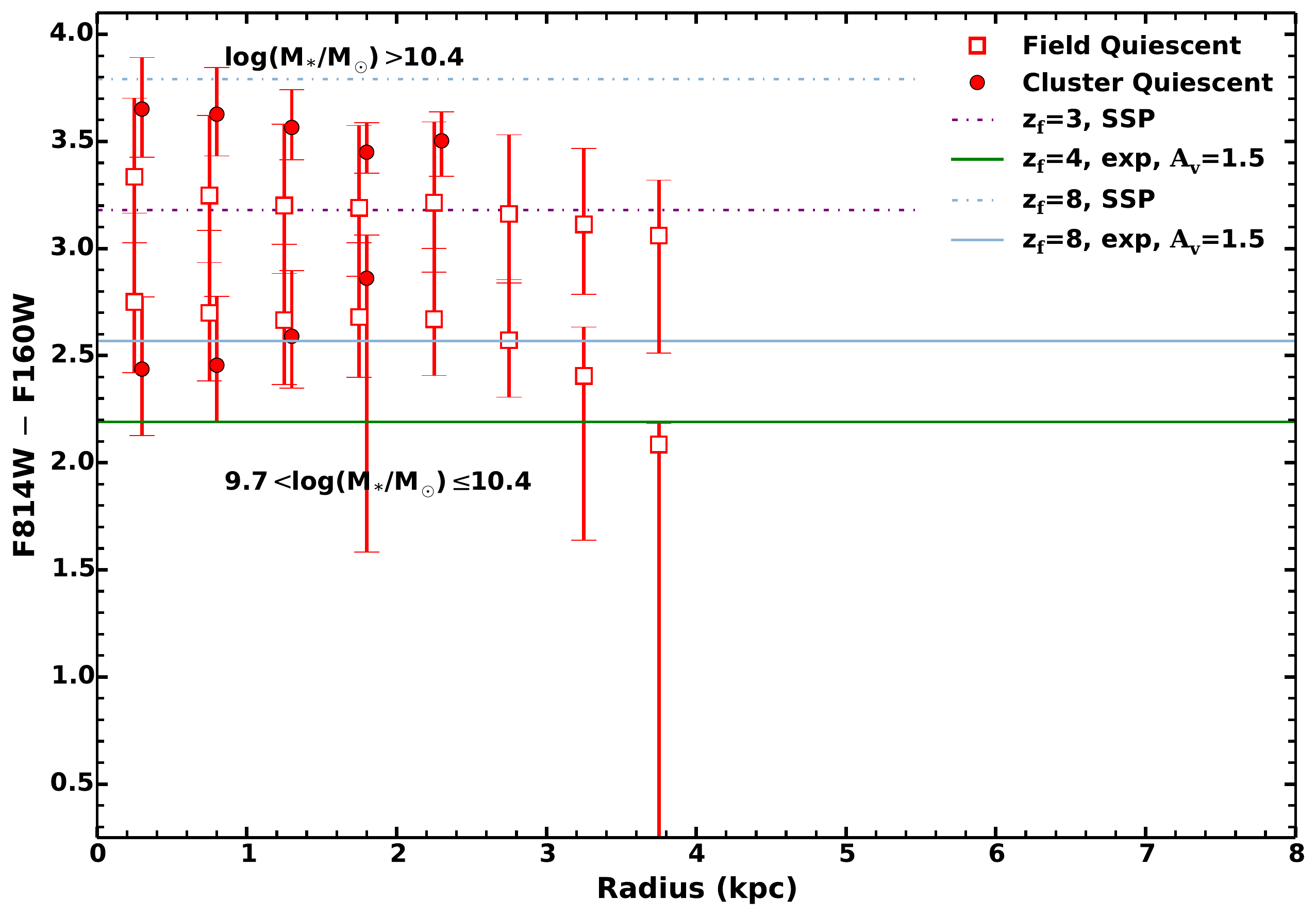}\\
  \epsscale{1}\plotone{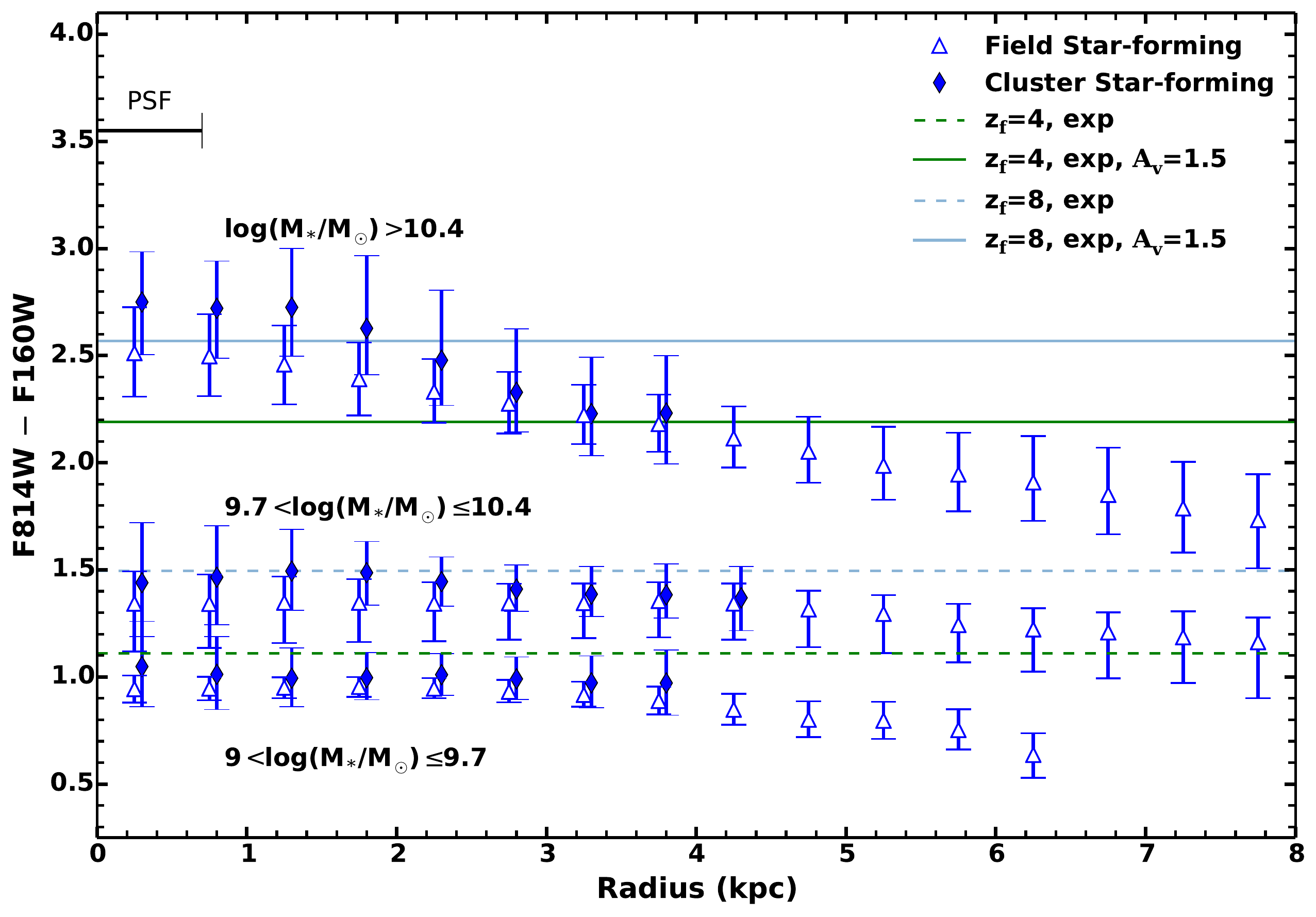}
   \caption{Observed radial color profilest for quiescent (top panel) and star-forming (bottom panel) field and cluster galaxies. Quiescent cluster (field) galaxies are shown as filled (open) red circles (squares). Star-forming cluster (field) galaxies are shown as filled (open) blue diamonds (triangles). In each panel our samples are separated by stellar mass. We show the extent of the HWHM of the {\HST}/F160W PSF as a black line. The dashed lines are colors calculated using stellar population evolution models from {\tt EZGAL}. Models with exponentially declining star-formation are labeled "exp" while models with a single stellar population are labeled "SSP". We find that stellar evolution models with no dust and solar metallicity are consistent with the low and intermediate mass star-forming sample. For the high mass star-forming sample, models with dust extinction of $A_{v}=1.5$ are necessary to account for the redder colors. Single stellar population models are consistent with the colors we find for our high mass quiescent galaxy sample. However, the colors of the intermediate mass quiescent sample are consistent with exponentially decreasing star formation and a dust extinction of $A_{v}=1.5$.}
 \label{fig:color}
\end{figure*}
\begin{table*}[]
\begin{center}
\caption{Mass-Color Relation (F814W$-$F160W) for Star-forming and Quiescent, Field and Cluster Galaxies}
\begin{tabular}{c c c c c c}
\hline\hline
&&Quiescent&&Star-forming \\[0.5ex]
\hline
Mass bin & Environment & \# of galaxies & F814W$-$F160W & \# of galaxies & F814W$-$F160W\\
\hline\\[-1.5ex]
$9\leq$~{\M}~$\leq9.7$ & Field & 1 & --- & 243 & $0.96\pm{0.02}$\\[0.5ex]
 & Cluster & 0 & --- & 23 & $1.27\pm{0.14}$ \\[0.5ex]
\\[-2.5ex]
&$\Delta_{FC}$$^{**}$& &---&&$\phantom{00000}-0.32\pm{0.14}$ ($2.30\sigma$) \\
 \hline\\
$9.7\textless$~{\M}~$\leq10.4$ & Field & 16 & $2.95\pm{0.16}$ & 136 & $1.64\pm{0.10}$\\[0.5ex] 
 & Cluster & 4 & $2.50\pm{0.15}$ & 18 & $1.93\pm{0.16}$\\[0.5ex]
\\[-2.5ex]
 &$\Delta_{FC}$& &$\phantom{000000)}0.45\pm{0.22}$ ($2.05\sigma$)&&$\phantom{00000}-0.29\pm{0.19}$ ($1.53\sigma$)\\
 \hline\\
{\M}~$\textgreater10.4$ & Field & 18 & $3.34\pm{0.12}$ & 64 & $2.41\pm{0.06}$\\[0.5ex] 
 & Cluster & 5 & $3.54\pm{0.08}$ & 9 & $2.77\pm{0.20}$\\[0.5ex]
 \\[-2.5ex]
 &$\Delta_{FC}$& &$\phantom{00000}-0.20\pm{0.14}$ ($1.39\sigma$)&&$\phantom{00000}-0.36\pm{0.21}$ ($1.71\sigma$)\\
\hline
\end{tabular}
\end{center}
\label{tab:nums}
\vglue -2ex
		$\phantom{000000000}$$^{**}$ $\Delta_{FC}\equiv$ Field~$-$~Cluster
\end{table*}

To reduce contamination from neighboring galaxies we create a mask for each galaxy thumbnail that flags all objects in the image except the central galaxy. 
The masking was accomplished by using SExtractor with a detection threshold of $1.2\sigma$ above the background rms level to create a bad pixel mask.  

To probe the global colors of our sample and how they vary with each sample, we use an aperture, D$=0.6''$, that contains a large fraction of the global flux for these galaxies.
We centre the aperture on the galaxy and measure the flux in both the {\HST}/F814W and {\HST}/F160W images.
Using the zero-point for each filter we convert the flux to AB magnitudes.
We show the observed F814W$-$F160W color, roughly equivalent to a rest-frame U$-$V color, versus mass relation in Figure \ref{fig:mvc}.

As seen in Figure \ref{fig:mvc}, galaxy colors are mass dependent and become redder as mass increases for both star-forming and quiescent galaxies \citep[see][]{2010ApJ...721..193P}.
To disentangle this effect, and see if there is an environmental dependence, we separate our field and cluster galaxies by mass and then calculate their average colors.
We defined our mass bins so that we have roughly equal numbers of quiescent galaxies in each bin.
The mass bins and observed colors for star-forming and quiescent, field and cluster galaxies, are listed in Table \ref{tab:nums}.
 
We see an evolution towards redder colors as a function of mass in the average colors of both quiescent and star-forming, field and cluster galaxies.
The mean color is $18\%$ redder for field quiescent galaxies with $9.7\textless$~{\M}~$\leq10.4$ than the mean color for cluster galaxies of the same mass. 
We find no significant difference in the mean color of field and cluster quiescent galaxies with {\M}~$\textgreater10.4$.

In each mass bin, cluster star-forming galaxies have colors that are $20\%$ redder than their field counter-parts.
A KS test indicates that the star-forming field and cluster color distributions differ by $3.64\sigma$.
The average colors for each environment are listed in Table 2.

\subsection{Radial Color Profiles}
In addition to individual galaxy colors, color gradients are an effective means of studying  a galaxies' radial distribution of stellar populations \citep[e.g.,][]{2012arXiv1202.0494W}.  
By comparing color gradients of field and cluster galaxies it is possible to see if environment plays a role in determining the stellar populations of galaxies \citep{vanDokkum2010}.

We utilized image stacking to create deep averaged images of our samples of field and cluster, star-forming and quiescent galaxies.
We use the same mass bins, image stamps and masks created to measure the individual galaxy colors for the image stacks (Table \ref{tab:nums}).
Since we are stacking images we also run SExtractor \citep{1996A&AS..117..393B} and use pixel-by-pixel interpolation shifts via IRAF's IMCOPY package to ensure that all galaxy centers coincide with the central pixel of the image thumbnail. 
Additionally, galaxy images are normalized by their K-band flux in our F814W and F160W stacks so that bright galaxies do not dominate.
The galaxy image thumbnails are averaged via IRAF's IMCOMBINE with a bad pixel flag that gives masked objects zero weight. 
We repeat this process for each environment-mass bin for the F814 {\HST}/ACS and F160W {\HST}/WFC3 images. 
The low-mass field star-forming image stacks are the deepest with an increased S/N of $\sim \sqrt{243}$. 

We then measured the azimuthially averaged radial light profiles for the image stacks in the two {\HST} images.
Radial light profiles from each image stack were measured by averaging pixels in radial bins using a custom python code. 
The difference of these radial light profiles is the observed color.

We show the observed radial color for the cluster and field galaxies in Figure \ref{fig:color} with $1\sigma$ errors derived from bootstrapping each sample 1000 times.
The solid and dashed lines are colors calculated using stellar population evolution models from {\tt EZGAL}\footnote{http://www.baryons.org/ezgal/model}.
The stellar population evolution models are based on \citet{BruzualCharlot2003} models.
We considered models with a single stellar population (SSP) and solar metallicity for the quiescent galaxies.
For the star-forming galaxies we used models with exponentially declining star-formation and $\tau=1$ Gyr, solar metallicity and dust extinction ranging between A$_{v}=0$ to $2.5$.

The radial color profiles of low mass field and cluster star-forming galaxies are consistent within their errors, flat, and extend to $\sim6$ kpc.
The observed colors for these galaxies are consistent with {\tt EZGAL} models which have no dust and $z_{f}\textless4$.

The intermediate mass star-forming field galaxies have deep profiles and we can trace their color out to $\sim8$ kpc.
However, there is no color gradient for these galaxies and their colors are equivalent with the cluster star-forming galaxies at the same mass. 
The colors of star-forming field and cluster intermediate mass galaxies correspond with {\tt EZGAL} models which have no dust and $z_{f}\textgreater4$.

The high mass field and cluster star-forming galaxies are consistent in color which is more red than for the lower mass galaxies.
The observed colors of star-forming field and cluster high mass galaxies match {\tt EZGAL} models which have dust attenuation of $A_{v}=1.5$ and $z_{f}\textgreater4$.
Both high mass field and cluster star-forming galaxies show a negative color gradient towards bluer colors at r~$\textgreater2$ kpc.

In the top panel of Figure \ref{fig:color}, we show that the intermediate mass field and cluster quiescent galaxies have profiles that are consistent within their errors.
The intermediate mass field quiescent galaxies may have bluer colors at larger radius, however we do not have a deep enough image stack (i.e. too few galaxies) to distinguish this.
The colors of these galaxies are interesting as they are not consistent with models containing simply an old SSP.
However, models with exponentially declining star-formation ($\tau=1$ Gyr), dust attenuation of $A_{v}=1.5$, and and $z_{f}\textgreater4$ have colors which match those of our sample.

The high mass quiescent field and cluster galaxies are consistent in color and have colors analogous with a SSP and $z_{f}\textgreater3$.
As for the intermediate mass field and cluster quiescent galaxies, the high mass field and cluster galaxy stacks are also shallow and do not extend to large enough radius to potentially reveal a significant color gradient. 

\section{Discussion}
\label{sec:dis}
For the first time, we have studied the relationship between environment and structural/stellar properties of star-forming and quiescent galaxies at $z=2.1$.
We found that at $z=2.1$, environment may be beginning to influence the sizes and stellar populations of star-forming galaxies.
However, at this epoch, it does not appear that environment is affecting the sizes or stellar populations of quiescent galaxies.
\subsection{Quiescent Galaxies}
Our sample of cluster quiescent galaxies is small and we suffer from poor statistics which drives the error in obtaining a robust comparison to our quiescent field sample.
The size difference measured between cluster and field galaxies, ${\Delta}r_{1/2, maj}$, has an error of 0.69{\kpc}, which is $32\%$ of the cluster galaxy size and $38\%$ of the field galaxy size.
Therefore, if the environment effects the sizes of quiescent galaxies at $\leq0.7$ kpc we would not be sensitive to it.
However, we do note that there is only one cluster quiescent galaxy with {\M}$\textgreater10.5$ and $r_{1/2, maj}\leq1${\kpc}, suggesting a lack of massive compact cluster quiescent galaxies.
This is in agreement with \citet{Papovich2012} who found a lack of massive compact cluster quiescent galaxies compared to the field at fixed mass. 

Mergers are thought to play a major role in the growth of massive galaxies \citep[e.g.,][]{2009ApJ...699L.178N}.
In higher density regions where clusters are still virializing, interactions between galaxies are more common and quiescent cluster galaxies could be undergoing mergers.
Slight differences in stellar populations, or colors, for field and cluster galaxies are a method of identifying growth via mergers.
By analysing the mass-color relation for individual field and cluster galaxies, separated by mass, we can see if there is a difference in color between the two environments.
The mean color of intermediate mass field quiescent galaxies is $18\%$ ($2.05\sigma$) redder than cluster galaxies at the same mass.
There is only a $6\%$ ($1.39\sigma$) difference in the mean color of high mass field and cluster quiescent galaxies. 
The lack of a significant color difference for quiescent cluster galaxies indicates that mergers are not yet occurring or we are not sensitive enough to detect them.
Another way to look for color differences is to use radial color profiles to distinguish if mass is being added to the galaxy.

The observed radial color profiles of our intermediate mass field and cluster quiescent galaxies only reach $\sim4${\kpc} and are consistent within their errors.
We find that the radial color profiles of the high mass quiescent sample are consistent across environment as well.
It is possible that both the intermediate and high mass quiescent galaxies are accreting mass at large radius and thus have bluer colors at large galactic radius, however, we are not sensitive enough to detect it.

At $z\textless2$ there is evidence that cluster quiescent galaxies are larger in size than coeval field quiescent galaxies, so this growth must occur over a short timescale \citep{Papovich2012,2013MNRAS.435..207L} or the difference in size at this epoch is too weak for us to detect.

\subsection{Star-forming Galaxies}
\begin{figure}[]
\centering
  \epsscale{1.1}\plotone{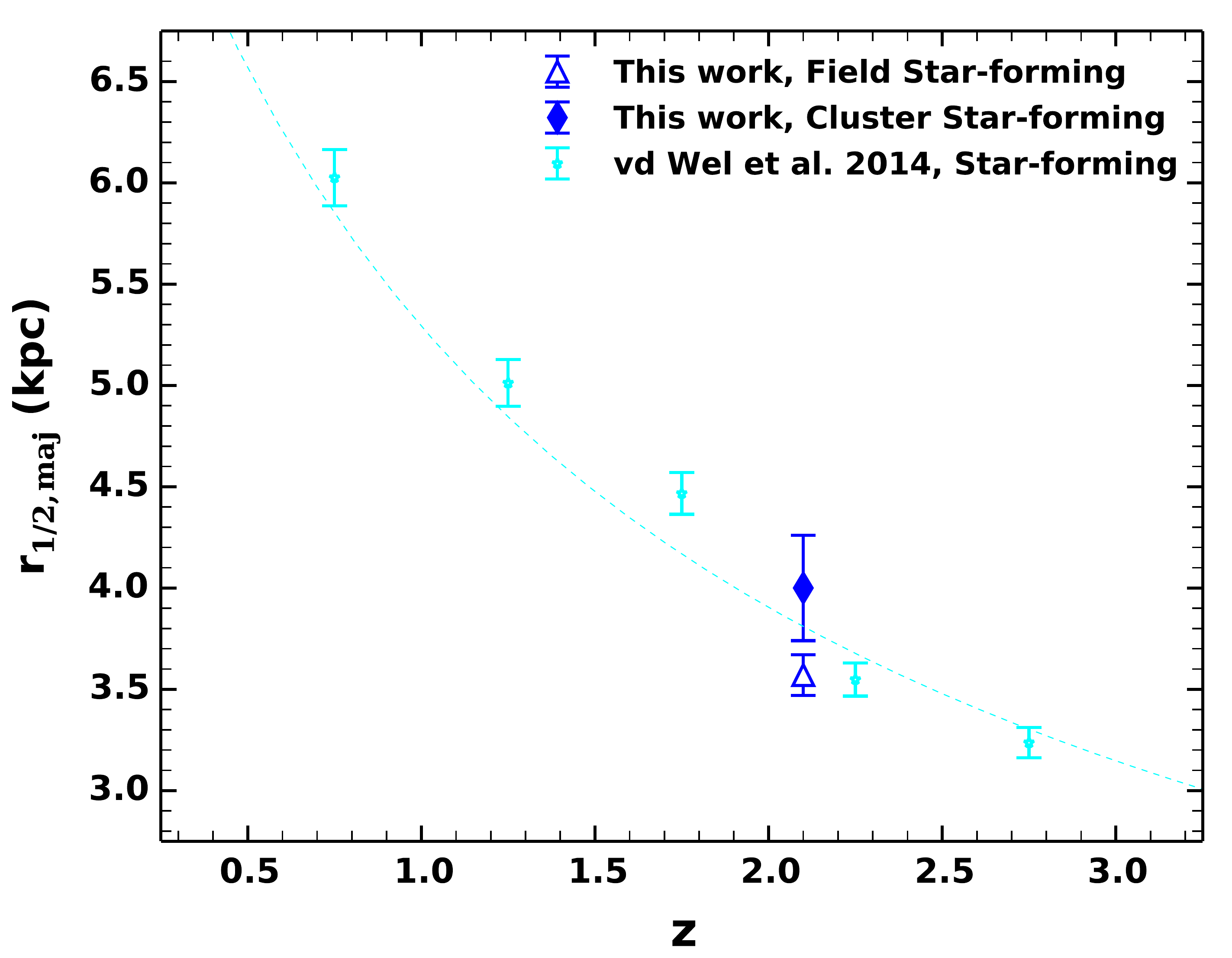}
     \caption{Evolution of the mass-size relation at fixed stellar mass of {\M}~$=10.7$. The large blue open and closed points are the best-fit sizes of field and cluster star-forming galaxies from this work. The small cyan points are the best-fit sizes of star-forming galaxies from \citep{2014ApJ...788...28V} for different redshifts. Their fitted relation for these sizes is shown as a cyan dashed line.}
 \label{fig:money}
\end{figure}

In Figure \ref{fig:money}, we show the evolution of the size mass relation for star-forming galaxies using the sizes and best fit relation of \citet{2014ApJ...788...28V}.
The mean size that we find for cluster galaxies is $12\%$ ($1.54\sigma$) larger than the mean size of field galaxies at the same mass.
In addition, performing a KS test indicates that the star-forming field and cluster size distributions differ at a significance of $2.10\sigma$.
Our mean size for cluster star-forming galaxies lies on the van der Wel et al. fitted relation, however, differs by $10\%$ from their mean size for star-forming galaxies at $z=2.25$.
The mean size of our field star-forming galaxies is consistent with the mean size \citet{2014ApJ...788...28V} found for star-forming galaxies at $z=2.25$, however, it does not lie on the relation at $z=2.1$.
\citet{2014ApJ...788...28V} do not differentiate between field/group/cluster galaxies in their sample selection and if environmental effects are not corrected for, then their average sizes would be larger than what would be found for a true field population.
The size difference we find between star-forming field and cluster galaxies is not consistent with \citet{2013ApJ...770...58B} who found no significant differences in the sizes of star-forming field and cluster galaxies at $z=1.6$.
Additionally, \citet{2013MNRAS.435..207L} found no environmental dependence for the mass-size relation of star-forming galaxies at $z=1-2$.

The fact that we find a significant difference in the mean sizes of field and cluster star-forming galaxies suggests that the cluster environment may be accelerating the evolution of massive star-forming galaxies.
\citet{vanDokkum2010} found that galaxies with {\M}~$\textgreater11.1$ grow preferentially via minor mergers from $0\leq{z}\leq2$.
At $z=2$, we do not have a significant number of galaxies above this mass limit, however, the star-forming galaxies in our sample with {\M}~$\textgreater10.4$ could also be growing via minor mergers.
Here, we analyze the stellar populations of our field and cluster star-forming galaxies samples, by using their observed colors, to look for signatures of minor mergers.

We find that the mean colors of star-forming cluster galaxies are $20\%$ redder than field galaxies at all masses.
After performing a KS test on the two color distributions we find that they differ by $3.64\sigma$.
This is suggestive that environment is beginning to influence the stellar populations of these galaxies.

The radial color profiles of star-forming galaxies can be used to distinguish if minor mergers are influencing their growth.  
We find that low and intermediate mass, field and cluster star-forming galaxies have color profiles that are consistent and flat.
We find that high mass field and cluster star-forming galaxies have bluer colors at radii above 2 kpc.
This is consistent with \citet{2011ApJ...735L..22S} who find negative color gradients for both star-forming and quiescent galaxies at $z\sim2$ with $10.1\leq${\M}$\leq11.1$. 
This is suggestive that both field and cluster star-forming galaxies are experiencing growth via minor mergers, however, we do not have the sensitivity to determine whether the color profiles at large radii of star-forming cluster galaxies become steeper than the profiles of star-forming field galaxies.
This would be important to quantify and to determine if minor mergers are more predominate in the cluster environment or if other mechanisms are causing the larger sizes for the star-forming cluster galaxies.

\section{Conclusions}
\label{sec:con}
Our aim was to determine the effects of environment on galaxy evolution using a galaxy cluster at $z=2.1$.
We created a sample of field and cluster galaxies with {\M}~$\geq9$ and used the UVJ rest-frame color-color diagram to separate them into star-forming and quiescent.
We utilized the morphological catalog of \citet{2014ApJ...788...28V} to analyse the size versus mass relation and distribution of {\n} indices for this sample of galaxies.
We further analyzed galaxy color gradients as a function of mass and environment.
Our main results are the following:

\begin{itemize}
  \item~We find that the mass normalized ({\M}$=10.7)$ sizes of cluster star-forming galaxies are $12\%$ larger, $1.5\sigma$, than field star-forming galaxies. A KS test shows that the distribution of sizes for field and cluster star-forming galaxies differs by $2.1\sigma$. However, the {\n} indices of these two populations are consistent within the errors.
  \item~Mean observed F814W-F160W colors for star-forming cluster galaxies are $20\%$ redder, than field galaxies at all masses. A KS test confirms that the color distributions of the two populations differ by $3.64\sigma$.
  \item~Radial observed F814W-F160W color profiles for star-forming field and cluster galaxies are consistent for each mass bin. A color negative gradient is observed in both field and cluster star-forming galaxies with {\M}$\textgreater10.4$, therefore, we cannot distinguish the source of the larger sizes of cluster star-forming galaxies. No color gradients are observed for field or cluster star-forming galaxies with {\M}$\leq10.4$.
  \item~Quiescent field and cluster galaxies are consistent in size and in {\n} index. However, we are only sensitive to differences of 0.7 kpc or greater due to our sample size.
  \item~Mean colors for quiescent field galaxies with $9.7\textless$~{\M}~$\leq10.4$ are $18\%$ redder, $2\sigma$, than cluster galaxies with the same mass. The mean colors are the same across environment for higher masses
   \item~Radial observed F814W-F160W color profiles for quiescent field and cluster galaxies are consistent for each mass bin and flat.
   \end{itemize}  
The combination of accurate photometric redshifts, catalogs of structural parameters, and image stacking has allowed us to probe a high redshift sample of field and cluster galaxies.
Our results imply that the effect of environment on galaxy sizes at $z=2.1$ is only significant for star-forming galaxies.
Even though there is evidence that our cluster is still in the early stages of formation \citep{Spitler2012}, we are able to detect a difference in the sizes and stellar populations of star-forming cluster galaxies compared to coeval field galaxies.
The negative color gradient of massive star-forming cluster galaxies suggests growth via minor mergers, although field galaxies at the same mass also display similar negative color gradients.
We require deeper imaging to determine if the negative color gradient for star-forming cluster galaxies extends as far as the field population.
At $z\textless2$ there is evidence that quiescent cluster galaxies are larger in size than coeval field galaxies so this growth must occur over a short timescale \citep{2013MNRAS.435..207L,Papovich2012}. 
The early stage of formation of our cluster could explain why we do not see larger sizes for cluster quiescent galaxies.
The mechanisms which affect star formation and general mass growth of galaxies in dense environments are poorly understood at high redshift.
To distinguish between which growth mechanisms are dominate, and how they evolve with time, more studies which use larger samples of cluster and field galaxies at $1\textless{z}\textless2$ are necessary.

\acknowledgments 
We would like to thank the referee for their constructive and helpful comments.
Research support to R.J.A is provided by the Australian Astronomical Observatory. 
GGK was supported by an Australian Research Council Future Fellowship FT140100933.
This work was supported by the National Science Foundation grant AST-1009707. 
L.R.S. acknowledges funding from a Australian Research Council (ARC) Discovery Program (DP) grant DP1094370 and Access to Major Research Facilities Program which is supported by the Commonwealth of Australia under the International Science Linkages program. 
Australian access to the Magellan Telescopes was supported through the National Collaborative Research Infrastructure Strategy of the Australian Federal Government. 
We thank the Las Companas Observatory for access to facilities for the ZFOURGE survey.

 \end{document}